\documentclass[journal]{IEEEtran}

\usepackage{cite}
\usepackage{hyperref}
\usepackage[pdftex]{graphicx}
\usepackage{epstopdf}
\usepackage{amsmath}
\usepackage{amssymb}
\usepackage{array}
\usepackage{wasysym}
\usepackage{bm}
\usepackage{units}
\usepackage[per-mode=symbol]{siunitx}
\usepackage{url}
\usepackage[usenames, dvipsnames]{color}
\usepackage{epstopdf}
\usepackage[ruled]{algorithm2e}
\usepackage{multirow}
\usepackage{booktabs}
\usepackage{lineno}
\usepackage{wasysym}
\usepackage{epstopdf}
\usepackage{nameref}

\usepackage{clipboard}

\begin{document}

\title{Investigations of a Robotic Testbed with Viscoelastic Liquid Cooled Actuators}

\author{Donghyun Kim,
        Junhyeok Ahn, Orion Campbell, Nicholas Paine, and Luis Sentis,        
        %
}


\maketitle

\begin{abstract}
We design, build, and thoroughly test a new type of actuator dubbed viscoelastic liquid cooled actuator (VLCA) for robotic applications. VLCAs excel in the following five critical axes of performance: energy efficiency, torque density, impact resistence, joint position and force controllability. We first study the design objectives and choices of the VLCA to enhance the performance on the needed criteria. We follow by an investigation on viscoelastic materials in terms of their damping, viscous and hysteresis properties as well as parameters related to the long-term performance. As part of the actuator design, we configure a disturbance observer to provide high-fidelity force control to enable a wide range of impedance control capabilities. We proceed to design a robotic system capable to lift payloads of $32.5~\si{\kilo\gram}$, which is three times larger than its own weight. In addition, we experiment with Cartesian trajectory control up to $2~\si{\hertz}$ with a vertical range of motion of $32~\si{\centi\meter}$ while carrying a payload of $10~\si{\kilo\gram}$. Finally, we perform experiments on impedance control and mechanical robustness by studying the response of the robotics testbed to hammering impacts and external force interactions.
\end{abstract}

\begin{IEEEkeywords}
Viscoelastic liquid cooled actuator, Torque feedback control, Impedance control.
\end{IEEEkeywords}

\IEEEpeerreviewmaketitle
\section{Introduction}
%
\IEEEPARstart{S}{eries} \Copy{revised_introduction}{elastic actuators (SEAs) \cite{Pratt:1995ky} have been extensively used in robotics~\cite{Henze:2016ct,Paine:2015wp} due to their impact resistance and high-fidelity torque controllability. One drawback of SEAs is the difficulty that arises when using a joint position controller due to the presence of the elastic element in the drivetrain. To remedy this problem the addition of dampers has been previously considered~\cite{hurst2004series, Kashiri:2015ct, CheeMengChew:de}. However, incorporating mechanical dampers makes actuators bulky and increases their mechanical complexity. 

\Copy{avoid_damper}{One way to avoid this complexity is to employ elastomers instead of metal springs. Using a viscoelastic material instead of combined spring-damper systems enables compactness~\cite{Rollinson:2014dd} and simplified drivetrains~\cite{Abe:2012kk}.} However, it is difficult to achieve high bandwidth torque control due to the nonlinear behavior of elastomers. To address this difficulty, \cite{Austin:2015gj} models the force-displacement curve of elastomer using a ``standard linear model.'' The estimated elastomer force is employed in a closed-loop force controller. Unfortunately, the hysteresis in the urethane elastomer destabilized the system at frequencies above 2 \si{\hertz}. In contrast our controllers achieve a bandwidth of 70 \si{\hertz}. The study on \cite{Rollinson:2013kj} accomplishes reasonably good torque control performance, but the range of torques is small to ensure that the elastomer operates in the linear region; our design and control methods described here achieve more than an order of magnitude higher range of torques with high fidelity tracking.

To sufficiently address the nonlinear behavior of elastomers, which severely reduce force control performance, we empirically analyze various viscoelastic materials with a custom-built elastomer testbed. We measure each material's linearity, creep, compression set, and damping under preloaded conditions, which is a study under-documented in the academic literature. To achieve stable and accurate force control, we study various feedback control schemes. In a previous work, we showed that the active passivity obtained from motor velocity feedback \cite{Kim:2016jg} and model-based control such as disturbance observer (DOB) \cite{Paine:fz} play an essential role in achieving high-fidelity force feedback control. Here, we analyze the phase margins of various feedback controllers and empirically show their operation in the new actuators. We verify the stability and accuracy of our controllers by studying impedance control and impact tests.

To test our new actuator, we have designed a two degree-of-freedom (DOF), robotic testbed, shown in Fig. \ref{fig:single_leg_testbed}. It integrates two of our new actuators, one in the ankle, and another in the knee, while restricting motions to the sagittal plane. With the foot bolted to the floor for initial tests, weight plates can be loaded on the hip joint to serve as an end-effector payload. We test operational space control to show stable and accurate operational space impedance behaviors. We perform dynamic motions with high payloads to showcase another important aspect of our system, which is its cooling system aimed at significantly increasing the power of the robot. 

The torque density of electric motors is often limited by sustainable core temperature. For this reason, the maximum continuous torque achieved by these motors can be significantly enhanced using an effective cooling system. Our previous study \cite{Paine:2015bs} analyzed the improvements on achievable power based on thermal data of electric motors and proposed metrics for design of cooling systems. Based on the metrics from that study, we chose a 120 \si{\watt} Maxon EC-max 40, which is expected to exert 3.59 times larger continuous torque when using the proposed liquid cooling system. We demonstrate the effectiveness of liquid cooling by exerting 860\si{\newton} continous force during 5 \si{\minute} and 4500\si{\newton} peak force during 0.5\si{\second} while keeping the core temperatures below 115\si{\degreeCelsius}, which is much smaller than the maximum, 155\si{\degreeCelsius}. We accurately track fast motions of 2 \si{\hertz} while carrying a 10 \si{\kilo\gram} payload for endurance tests. In addition we perform heavy lift tests with a payload of 32.5 \si{\kilo\gram} keeping the motor temperatures under 80\si{\degreeCelsius}.


\Copy{contribution}{The main contribution of this paper is the introduction of a new viscoelastic liquid cooled actuator and a thorough study of its performance and its use on a multidof testbed. We demonstrate that the use of liquid cooling and the elastomer significantly improve joint position controllability and power density over traditional SEAs. More concretely, we 1) design a new actuator, dubbed the VLCA, 2) extensively study viscoelastic materials, 3) extensively analyze torque feedback controllers for VLCAs, and 4) examine the performance in a multidof prototype.}
}
\section{Background}
\label{sec:background}
Existing actuators can be characterized using four criteria: power source (electric or hydraulic), cooling type (air or liquid), elasticity of the drivetrain (rigid or elastic), and drivetrain type (direct, harmonic drive, ball screw, etc.) \cite{hunter1991comparative, Paine:2014vc}. One of the most powerful and common solutions is the combination of hydraulic, liquid-cooling, rigid and direct drive actuation. This achieves high power-to-weight and torque-to-weight ratios, joint position controllability, and shock tolerance. Existing robots that use this type of actuators include Atlas, Spot, Big Dog, and Wildcat of Boston Dynamics, BLEEX of Berkeley \cite{zoss2006biomechanical}, and HyQ of IIT \cite{semini2010hyq}. However, hydraulics are less energy efficient primarily because they require more energy transformations \cite{bhounsule2014low}. Typically, a gasoline engine or electric motor spins a pump, which compresses hydraulic fluid, which is modulated by a hydraulic servo valve, which finally causes a hydraulic piston to apply a force. Each stage in this process incurs some efficiency loss, and the total losses can be very significant.

The combination of electric, air-cooled, rigid, and harmonic drive actuators are other widely used actuation types. Some robots utilizing these actuator types include Asimo of Honda, HRP2,3,4 of AIST \cite{kanehira2002design}, HUBO of KAIST \cite{park2005mechanical}, REEM-C of PAL Robotics, JOHNNIE and LOLA of Tech. Univ. of Munich \cite{gienger2001towards, lohmeier2006modular}, CHIMP of CMU \cite{stentz2015chimp}, Robosimian of NASA JPL \cite{karumanchi2017team}, and more. These actuators have precise position control and high torque density. \Copy{torque_density}{For example, LOLA's theoretical knee peak torque-density ($129 \si{\newton\meter\per\kilo\gram}$) is comparable to ours ($107 \si{\newton\meter\per\kilo\gram}$), although they did not validate their number experimentally and their max speed is roughly 2/3 of our max speed \cite{lohmeier2006modular}. Compared to us, low shock tolerance, low fidelity force sensing, and low efficiency gearboxes are common drawbacks of these type of actuators. According to Harmonic Drive AG’s catalog, the efficiency of harmonic drives may be as poor as 25\% and only increases above 80\% when optimal combinations of input shaft speed, ambient temperature, gear ratio, and lubrication are present. Conversely, the efficiency of our VLCA is consistently above 80\% due to the use of a ball screw mechanism.}

\cite{Urata:2010iw} used liquid cooling for electric, rigid, harmonic drive actuators to enhance continuous power-to-weight ratio. The robots using this type of actuation include SCHAFT and Jaxon~\cite{Kojima:2015ht}. These actuators share the advantages and disadvantages of electric, rigid, harmonic drive actuators, but have a significant increase of the continuous power output and torque density. One of our studies \cite{Paine:2015bs}, indicates a 2x increase in sustained power output by retrofitting an electric motor with liquid cooling. Other published results indicate a 6x increase in torque density through liquid cooling \cite{hunter1991comparative, aghili2007modular}, though such performance required custom-designing a motor specifically for liquid cooling. In our case we use an off-the-shelf electric motor. In contrast with our design, these actuators do not employ viscoelastic materials reducing their mechanical robustness and high quality force sensing and control.

Although the increased power density achieved via liquid cooling amplifies an electric actuator's power, the rigid drivetrain is still vulnerable to external impacts. To increase impact tolerance, many robots (e.g. Walkman and COMAN of IIT \cite{tsagarakis2013compliant}, Valkyrie of NASA \cite{radford2015valkyrie}, MABEL and MARLO in UMich \cite{grizzle2009mabel, ramezani2013feedback}, and StarlETH of ETH \cite{hutter2012starleth}) adopt electric, air-cooled, elastic, harmonic drive actuators. This type of actuation provides high quality force sensing, force control, impact resistance, and energy efficiency. However, precise joint position control is difficult because of the elasticity in the drivetrain and the coupled effect of force feedback control and realtime latencies \cite{Zhao:2015}. Low efficiency originating from the harmonic drives is another drawback. 

As an alternative to harmonic drives, ball screws are great drives for mechanical power transmission. SAFFiR, THOR, and ESCHER of Virginia Tech \cite{ lahr2013early, Lee:2014vs, knabe2015design}, M2V2 of IHMC \cite{pratt2008design}, Spring Flamingo of MIT \cite{pratt2000exploiting}, Hume of UT Austin \cite{Kim:2016jg}, and the X1 Mina exoskeleton of NASA \cite{rea2013x1} use electric, air-cooled, elastic, ball-screw drives. These actuators show energy efficiency, good power and force density, low noise force sensing, high fidelity force controllability, and low backlash. Compared to these actuators our design significantly reduces the bulk of the actuator and increases its joint position controllability. There are some other actuators that have special features such as the electric actuators used in MIT's cheetah \cite{seok2015design}, which allow for shock resistance through a transparent but backlash-prone drivetrain. However, the lack of passive damping limits the joint position controllability of these type of actuators compared to us. 

\section{viscoelastic material characterization}
\label{sec:visco}

\begin{figure*}
\centering
\includegraphics[width=2.0\columnwidth]{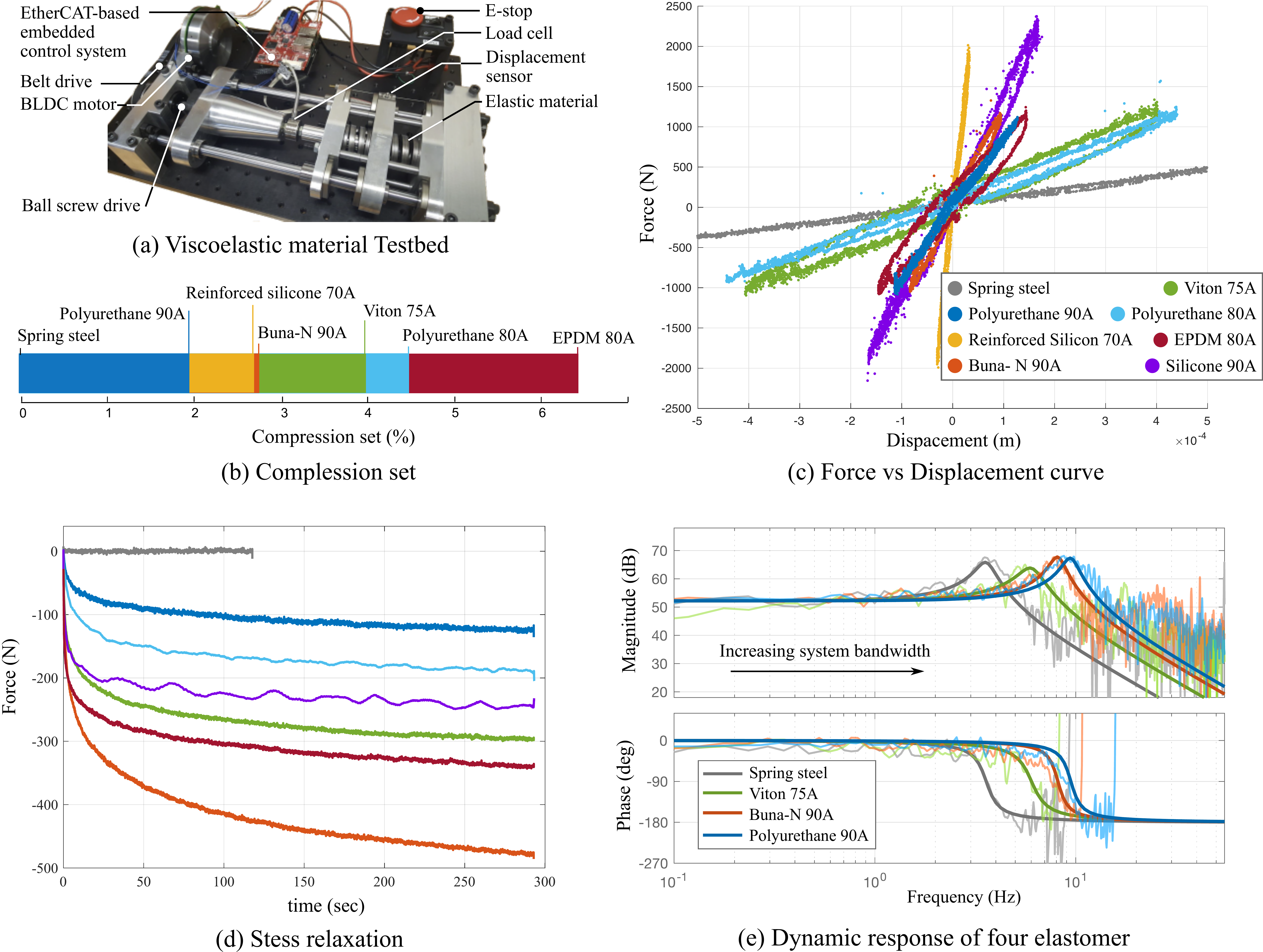}
\caption{{\bf Viscoelastic material test.} (a) The elastomer testbed is designed and constructed to study various material properties of candidate viscoelastic materials. (b) We measured each elastomers free length both before and after they were placed in the preloaded testbed. (c) A strong correlation between material hardness and the material’s stiffness can be observed. An exception to this correlation is the fabric reinforced silicone which we hypothesize had increased stiffness due to the inelastic nature of its reinforcing fabric. Nonlinear effects such as hysteresis can also be observed in this plot. (d) We command a rapid change in material displacements and then measured the material’s force change versus time for 300 seconds. Note that the test of reinforced silicone 70A is omitted due to its excessive stiffness. (d) Although the bandwidths of the four responses are different, their damping ratios (signal peak value) are relatively constant, which implies different damping.}
\label{fig:elastomer_test}
\vspace{-1.5mm}
\end{figure*}

The primary driver for using elastomers instead of metal springs is to benefit from their intrinsic damping properties. However, the mechanical properties of viscoelastic materials can be difficult to predict, thus making the design of an actuator based on these materials a challenging endeavor.

The most challenging aspect of incorporating elastomers into the structural path of an actuator is in estimating or modeling their  complex mechanical properties. Elastomers possess both hysteresis and strain-dependent stress, which result in nonlinear force displacement characteristics. Additionally, elastomers also exhibit time-varying stress-relaxation effects when exposed to a constant load. The result of this effect is a gradual reduction of restoration forces when operating under a load. A third challenge when using elastomers in compression is compression set. This phenomenon occurs when elastomers are subjected to compressive loads over long periods of time. An elastomer that has been compressed will exhibit a shorter free-length than an uncompressed elastomer. Compression set is a common failure mode for o-rings, and in our application, it could lead to actuator backlash if not accounted for properly.

To address these various engineering challenges we designed experiments to empirically measure the following four properties of our viscoelastic springs: 1) force versus displacement, 2) stress relaxation, 3) compression set, and 4) frequency response, which will be used to characterize each material's effective viscous damping. We built a viscoelastic material testbed, depicted in Fig.~\ref{fig:elastomer_test}(a), to measure each of these properties. We selected and tested the seven candidate materials that are listed in Table~\ref{tb:visco_results}. \Copy{material_dimension}{The dimension of the tested materials are fairly regular, with 46\si{\milli\meter} diameter and 27\si{\milli\meter} thickness.}
%
\begin{table*}
\centering
\begin{tabular}{>{\centering}m{0.35\columnwidth}  %
                >{\centering}m{0.13\columnwidth} %
                >{\centering}m{0.15\columnwidth} %
                >{\centering}m{0.2\columnwidth} %
                >{\centering}m{0.25\columnwidth} %
                >{\centering}m{0.25\columnwidth} %
                >{\centering}m{0.12\columnwidth} %
                >{\centering}m{0.15\columnwidth} @{}m{0pt}@{}}
\specialrule{1.5pt}{1pt}{1pt}
\vspace{1mm}
{Materials}
& {Compression set (\%) }
& {Linearity (R-square)}
& {Linear stiffness (\si{\newton\per\milli\meter})}
& {Preloaded elastic modulus (\si{\newton\per\milli\square\meter})}
& {Material damping (\si{\newton\second\per\meter})}
& {Creep (\%)}
& {Material Cost (\$)}
&\\[2.0mm] 
\hline
\hline
Spring steel & 0 & 0.996 & 860.8 & ~ & 0 & 0& - &\\ [1mm] \hline
Polyurethane 90A & 2 & 0.992 & 8109 & 112.5 & 16000 & 15.3 & 19.40 &\\ [1mm] \hline
Reinforced silicone 70A & 2.7 & 0.978 & 57570 & 798.7 & 242000 & - & 29.08 &\\ [1mm] \hline
Buna-N 90A & 2.8 & 0.975 & 11270 & 156.4 & 29000 & 25 & 51.47 &\\ [1mm] \hline
Viton 75A & 4 & 0.963 & 2430 & 33.7 & 9000 & 30.14 & 105.62 &\\ [1mm]
\hline
Polyurethane 80A & 4.5 & 0.993 & 2266 & 31.4 & 4000 & 16.8 & 19.40 &\\ [1mm] \hline 
EPDM 80A& 6.48 & 0.939 & 6499 & 90.2 & 16000 & 23.4 & 35.28 &\\ [1mm] \hline
Silicone 90A & - & 0.983 & 12460 & 172.9 & 37000 & 10.7 & 29.41 &\\ [1mm] \hline
\hspace{0.5mm}
\end{tabular}
\caption{Summary of viscoelastic materials}
\label{tb:visco_results}
\vspace{-6mm}
\end{table*}

\subsection{Compression set}
Compression set is the reduction in length of an elastomer after prolonged compression. The drawback of using materials with compression set in compliant actuation is that the materials must be installed with larger amounts of preload forces to avoid the material sliding out of place during usage. To measure this property, we measured each elastomer’s free length both before and after the elastomer was placed in the preloaded testbed. The result of our compression set experiments are summarized in Table~\ref{tb:visco_results}.

\subsection{Force versus displacement}
In the design of compliant actuation, it is essential to know how much a spring will compress given an applied force. This displacement determines the required sensitivity of a spring-deflection sensor and also affects mechanical aspects of the actuator such as usable actuator range of motion and clearance to other components due to Poisson ratio expansion. In this experiment, we identify the force versus displacement curves for the various elastomer springs. Experimental data for all eight springs as shown in Fig~\ref{fig:elastomer_test}(b). Note that there is a disagreement between our empirical measurements and the analytic model relating stiffness to hardness, i.e. the Gent's relation shown in \cite{Pucci:2002ev}. This mismatch arises because in our experiments the materials are preloaded whereas the analytical models assume unloaded materials.

\subsection{Stress relaxation}
Stress-relaxation is an undesirable property in compliant actuators for two reasons. First, the time-varying force degrades the quality of the compliant material as a force sensor. When a material with significant stress-relaxation properties is used, the only way to accurately estimate actuator force based on deflection data is to model the effect and then pass deflection data through this model to obtain a force estimate. This model introduces complexity and more room for error. The second reason stress-relaxation can be problematic is that it can lead to the loss of contact forces in compression-based spring structures. 

The experiment for stress relaxation is conducted as follow: 1) enforce a desired displacement to a material, 2) record the force data over time from the load cell, 3) subtract the initially measured force from all of the force data. Empirically measured stress-relaxation properties for each of the materials are shown in Fig.~\ref{fig:elastomer_test} (c), which represents force offsets as time goes under the same displacement enforced. Note that each material shows different initial force due to the different stiffness and each initial force data is subtracted in the plot.

\subsection{Dynamic response}
In regards to compliant actuation, the primary benefit of using an elastomer spring is its viscous properties, which can characterize the dynamic response of an actuator in series with such a component. To perform this experiment, we generate motor current to track an exponential chirp signal, testing frequencies between 0.001Hz and 200Hz. Given the input-output relation of the system, we can fit a second order transfer function to the experimental data to obtain an estimate of the system's viscous properties. \Copy{testbed_damping}{However, this measure also includes the viscoelastic testbed's ballscrew drive train friction (Fig.~\ref{fig:elastomer_test}(a)). To quantify the elastomer spring damping independently of the damping of the testbed drive train, the latter (8000 \si[per-mode=symbol]
{\newton\second\per\metre}) was first characterized using a metal spring, and then subtracted from subsequent tests of the elastomer springs to obtain estimates for the viscous properties of the elastomer materials.} Fig.~\ref{fig:elastomer_test}(d) shows the frequency response results for current input and force output of three different springs, while controlling the damping ratio. \Copy{elastomer_bandwidth}{The elastomers have higher stiffness than the metal spring, hence their natural frequencies are higher.} 

\subsection{Selection of Polyurethane 90A}
A variety of other experiments were conducted to strengthen our analysis and are summarized in Table~\ref{tb:visco_results}. Based on these results, Polyurethane 90A appears to be a strong candidate for viscoelastic actuators based on its high linearity (0.992), low compression set (2\%), low creep (15\%), and reasonably high damping (16000 \si{\newton\second\per\meter}). It is also the cheapest of the materials and comes in the largest variety of hardnesses and sizes.

\section{Viscoelastic Liquid Cooled Actuation}
\label{sec:vlca}
\begin{figure*}
\centering
\includegraphics[width=2.0\columnwidth]{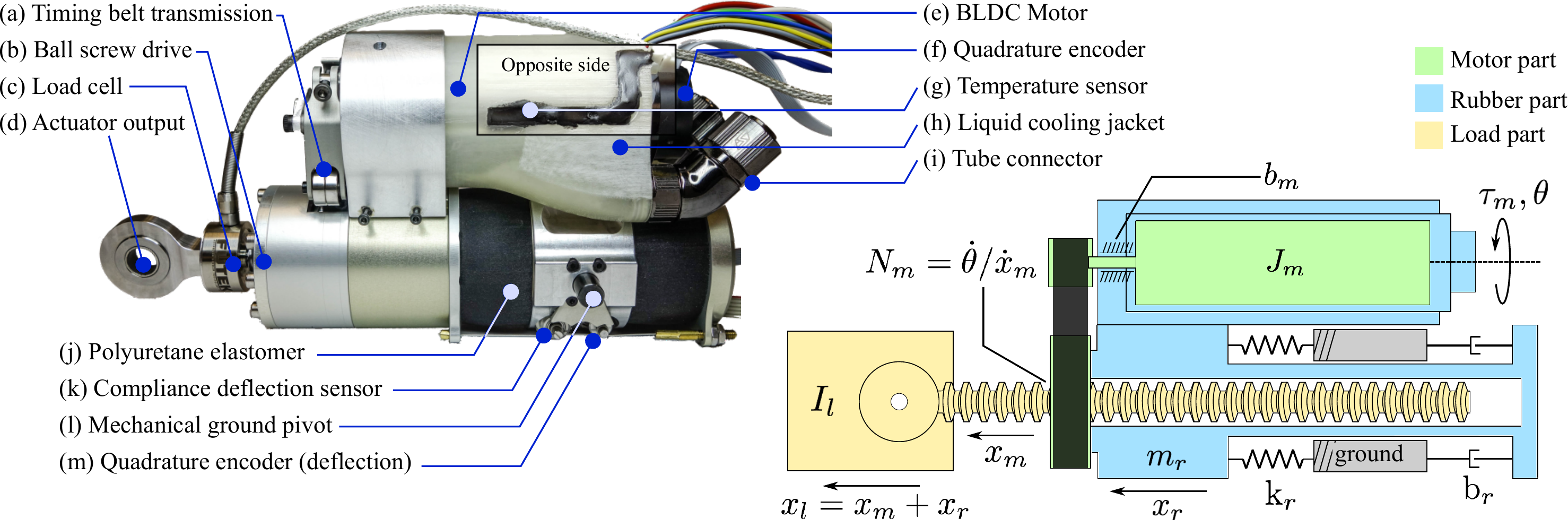}
\caption{{\bf Viscoelastic Liquid Cooled Actuator.} The labels are explanatory. In addition, the actuator contains five sensors: a load cell, a quadrature encoder for the electric motor, a temperature sensor, and two elastomer deflection sensors. One of the elastomer deflection sensors is absolute and the other one is a quadrature encoder. The quadrature encoder gives high quality velocity data of the elastomer deflection.}
\label{fig:actuator}
\vspace{-1.5mm}
\end{figure*}

The design objectives of the VLCA are 1) power density, 2) efficiency, 3) impact tolerance, 4) joint position controllability, and 5) force controllability. \Copy{avoid_damper2}{Compactness of actuators is also one of the critical design parameters, which encourage us to use elastomers instead of metal springs and mechanical dampers.}
Our previous work \cite{Paine:2015bs} shows a significant improvement in motor current, torque, output power and system efficiency for liquid cooled commercial off-the-shelf (COTS) electric motors and studied several Maxon motors for comparison. \Copy{thermal-blah}{As an extension of this previous work, in this new study we studied COTS motors and their thermal behavior models and selected the Maxon EC-max 40 brushless 120 $\si{\watt}$ (Fig. \ref{fig:actuator}(e)), with a custom housing designed for the liquid cooling system (Fig.~\ref{fig:actuator}(h)). The limit of continuous current increases by a factor of 3.59 when liquid convection is used for cooling the motor. Therefore, a continuous motor torque of 0.701 $\si[inter-unit-product={}\cdot{}]{\newton\metre}$ is theoretically achievable. Energetically, this actuator is designed to achieve 366 \si{\watt} continuous power and 1098\si{\watt} short-term power output with an 85\% ball screw efficiency (Fig. \ref{fig:actuator}(b)) since short-term power is generally three time larger than continuous power. With the total actuator mass of 1.692 \si{\kilo\gram}, this translates into a continuous power of 216\si{\watt\per \kilo\gram} and a short-term power of 650\si{\watt\per\kilo\gram}}. \Copy{mass_ignorance}{The liquid pump, radiator, and reservoir are products of Swiftech which weight approximately 1\si{\kilo\gram}}. By combining convection liquid cooling, high power brushless DC (BLDC) motors, and a high-efficiency ball screw, we aim to surpass existing electric actuation technologies with COTS motors in terms of power density.

In terms of controls, a common problem with conventional SEAs is their lack of physical damping at their mechanical output. As a result, active damping must be provided from torque produced by the motor \cite{Hutter:2011to}. However, the presence of signal latency and derivative signal filtering limit the amount by which this active damping can be increased, resulting in SEA driven robots achieving only relatively low output impedances \cite{Zhao:2015} and thus operating with limited joint position control accuracy and bandwidth. Our VLCA design incorporates damping directly into the compliant element itself, reducing the requirements placed on active damping efforts from the controller. The incorporation of passive damping aims to increase the output impedance while retaining compliance properties, resulting in higher joint position control bandwidth. The material properties we took into consideration will be introduced in Section \ref{sec:visco}. 
The retention of a compliant element in the VLCA drive enables the measurement of actuator forces based on deflection. The inclusion of a load cell (Fig. \ref{fig:actuator}(c)) on the actuator’s output serves as a redundant force sensor and is used to calibrate the force displacement characteristics of the viscoelastic element.

Mechanical power is transmitted when the motor turns a ball nut via a low-loss timing belt and pulley (Fig.~\ref{fig:actuator} (a)), which causes a ball screw to apply a force to the actuator's output (Fig.~\ref{fig:actuator}(d)). The rigid assembly consisting of the motor, ball screw, and ball nut connects in series to a compliant viscoelastic element (Fig.~\ref{fig:actuator}(j)), which connects to the mechanical ground of the actuator (Fig.~\ref{fig:actuator}(k)). When the actuator applies a force, the reaction force compresses the viscoelastic element. The viscoelastic element enables the actuator to be more shock tolerant than rigid actuators yet also maintain high output impedance due to the inherent damping in the elastomer.

\section{Actuator Force Feedback Control}
\label{sec:act_force_ctrl}

To demonstrate various impedance behaviors in operational space, robots must have a stable force controller. Stable and accurate operational space control (OSC) is not trivial to achieve because of the bandwidth interference between outer position feedback control (OSC) and inner torque feedback control \cite{Kim:2016jg}. Since stable torque control is a critical component for a successful OSC implementation, we extensively study various force feedback controls.

\begin{figure}
\centering
\includegraphics[width=1.0\columnwidth]{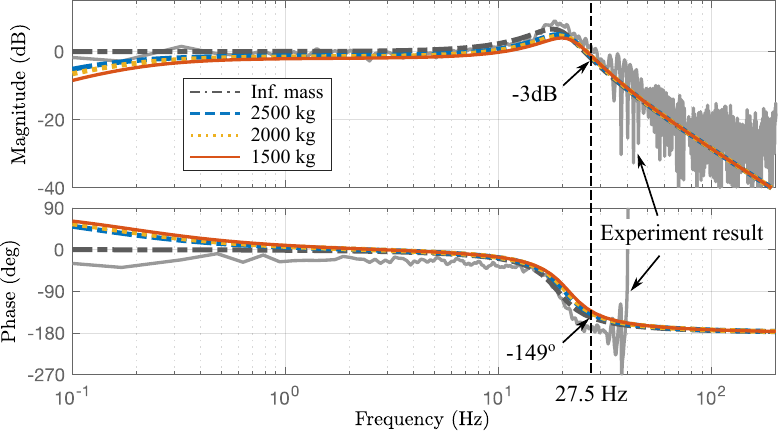}
\caption{{\bf Frequency response of VLCA.} Gray solid lines are experimental data and the other lines are estimated response with the model using empirically parameters.}
\label{fig:vlca_sys}
\vspace{-2mm}
\end{figure}

The first step in this analysis is to identify the actuator dynamics. The transfer functions of the reaction force sensed in the series elastic actuators (elastomer deflection) are well explained in \cite{YongsuPark:2016fo}. 
When the actuator output is fixed, the transfer function from the motor current input to the elastomer deflection is given by 
\begin{equation}
P_x = \frac{x_r}{i_m} = \frac{\eta k_{\tau} N_m}{(J_m N_m^2 + m_r)s^2 + (b_m N_m^2 + {\rm b}_r)s + {\rm k}_r},
\end{equation}
where $\eta$, $k_{\tau}$, $N_m$, and $i_m$ are the ball screw efficiency, the torque constant of a motor, the speed reduction ratio of the motor to the ball screw, and the current input for the motor, respectively.
The equations follow the nomenclature in Fig.~\ref{fig:vlca_sys}(a). We can find $\eta$, $k_{\tau}$, and $N_m$ in data sheets, which are 0.9, 0.0448 \si[inter-unit-product={}\cdot{}]{\newton\meter\per\ampere}, and 3316 respectively. \Copy{gear_ratio}{The gear ratio of the drivetrain is computed by dividing the speed reduction of pulleys (2.111) with lead length of the ball screw (0.004\si{\meter}) using the equation $2\pi\times 2.111 / 0.004$.} 

However, we need to experimentally identify ${\rm k}_r$, ${\rm b}_r$, $J_m$, and $b_m$. We infer ${\rm k}_r$ by dividing the force measurement from the load cell by the elastomer deflection. The other parameters are estimated by comparing the frequency response of the model and experimental data.
The frequency response test is done with the ankle actuator while prohibiting joint movement with a load and an offset force command. The results are presented in Fig.~\ref{fig:vlca_sys} with solid gray lines. \Copy{transfer_fn_input_output}{Note that the dotted gray lines are the estimated response from the transfer function (measured elastomer force/ input motor force) using the parameters of Table.~\ref{tb:vlca_id}.} The estimated response and experimental result match closely with one another, implying that the parameters we found are close to the actual values. 

\begin{table}
\centering
\begin{tabular}{>{\centering}m{0.17\columnwidth} %
                >{\centering}m{0.17\columnwidth} %
                >{\centering}m{0.12\columnwidth} %
                >{\centering}m{0.15\columnwidth} %
                >{\centering}m{0.16\columnwidth} 
                @{}m{0pt}@{}}
\specialrule{1.5pt}{1pt}{1pt}
{$J_m (\si{\kilo\gram\square\meter})$}
& {$b_m (\si{\newton\meter\second})$}
& {$m_r (\si{\kilo\gram})$}
& {${\rm b}_r (\si{\newton\second\per\meter})$}
& {${\rm k}_r (\si{\newton\per\meter})$}
&\\[1mm] 
\hline
\hline 
\num[output-exponent-marker = \text{e}]{3.8e-5} & 
\num[output-exponent-marker = \text{e}]{2.0e-4} & 1.3 & \num[output-exponent-marker = \text{e}]{2.0e4} & 5.5e6 &\\[1.5mm]
\hline
\vspace{2mm}
\end{tabular}
\caption{Actuator Parameters}
\label{tb:vlca_id}
\vspace{-6mm}
\end{table}
We also study the frequency response for different load masses to understand how the dynamics changes as the joint moves. When 10\si{\kilo\gram} is attached to the end of link, the reflected mass to the actuator varies from 1500\si{\kilo\gram} to 2500\si{\kilo\gram} because the length of the effective moment arm changes depending on joint position. In Fig.~\ref{fig:vlca_sys}(b), the bode plots are presented and the response is not significantly different than the fixed output case. Therefore, we design and analyze the feedback controller based on the fixed output dynamics.

\begin{figure}
\includegraphics[width=\columnwidth]{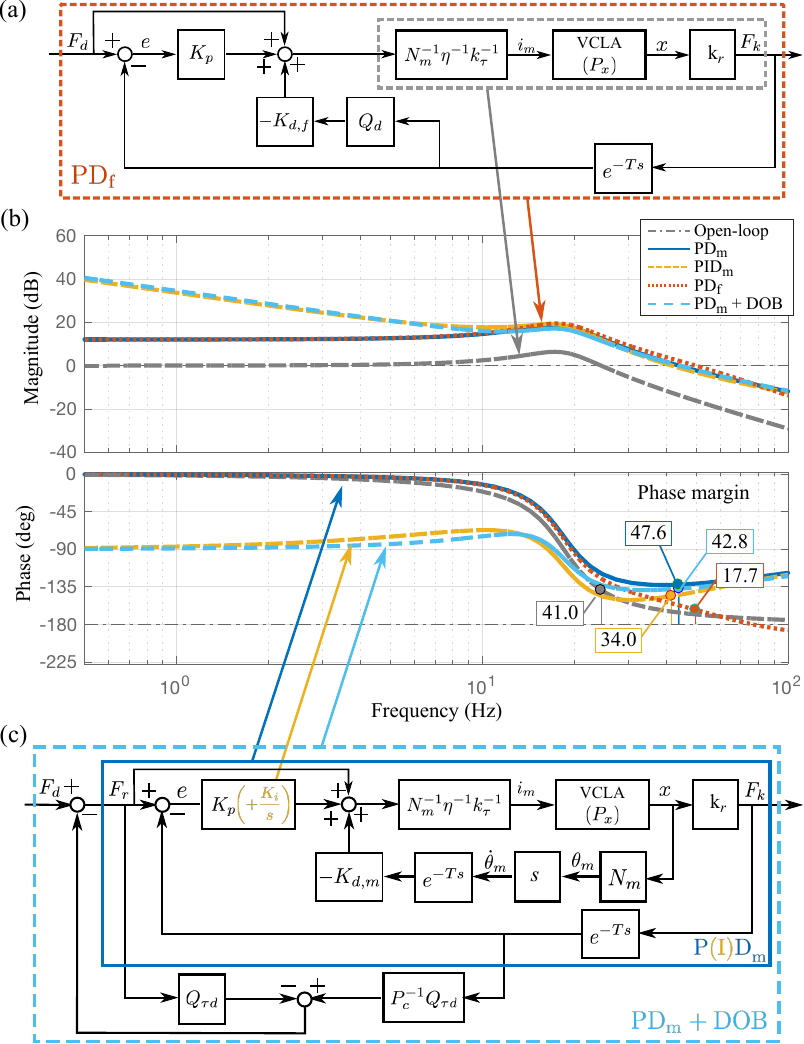}
\caption{{\bf Stability analysis of controllers.} Phase margins of each controllers and open-loop system are presented.}
\label{fig:stability_analysis}
\vspace{-2mm}
\end{figure}

For the force feedback controller, we first compare two options, which we have used in our previous studies \cite{Paine:fz,Kim:2016jg}:
\Copy{PD_f_PD_m}{
\begin{enumerate}
\item Proportional ($\rm{P}$) + Derivative ($\rm{D}_f$) using velocity signal obtained by a low-pass derivative filtered elastomer deflection
\item Proportional ($\rm{P}$) + Derivative ($\rm{D}_m$) using motor velocity signal measured by a quadrature encoder connected to a motor axis
\end{enumerate}
}
The second controller ($\rm{PD}_m$) has benefits over the first one ($\rm{PD}_f$) with respect to sensor signal quality. The velocity of motor is directly measured by a quadrature encoder rather than low-pass filtered elastomer deflection data, which is relatively noisy and lagged. In addition, Fig.~\ref{fig:stability_analysis} shows that the phase margin of the second controller (47.6)  is larger than the first one (17.1).

To remove the force tracking error at low frequencies, we consider two options: augmenting the controller either with integral control or with a DOB on the $\rm{PD}_m$ controller. To compare the two controllers, we analyzed the phase margins of all the mentioned controllers. First, we chose to focus on the location where the sensor data returns in order to address the time delay of digital controllers (Fig.~\ref{fig:stability_analysis} (a) and (c)). Next, we have to compute the open-loop transfer function for each closed loop system. For example, the $\rm{PD}_f$ controller's closed loop transfer function is \Copy{F_k_F_r}{ 
\begin{equation}
\label{eq:pd_f_tran_fn}
F_k = \frac{{\rm k}_r P_x}{N}\left( k_p (F_r - e^{-Ts} F_k) +F_r- k_{d,f} Q_d e^{-Ts} F_k \right),
\end{equation}
where $F_k$, $F_r$, $T$, and $Q_d$ are the measured force from a elastomer deflection, a reference force, a time delay, a low pass derivative filter, respectively.}For convenience, we use $N$ instead of the multiplication of three terms, $\eta k_{\tau} N_m$. When gathering the term with $e^{-Ts}$ of Eq.~\eqref{eq:pd_f_tran_fn}, we obtain 
\begin{equation}
\frac{F_k}{F_r} = \frac{{\rm k}_r P_x(K_p+1)/N}{1 + e^{-Ts}{\rm k}_r P_x (K_p + K_{d,f} Q_d)/N}.
\end{equation}
Then, the open-loop transfer function of the closed system with the time delay is 
\begin{equation}
P_{\rm{PD}_f}^{open} = {\rm k}_r P_x(K_p + K_{d,f}Q_d)/N.
\end{equation}
We can apply the same method for the $\rm{PID}_m$ and $\rm{PD}_m + \rm{DOB}$ controllers. 

The transfer function of $\rm{PID}_m$, which is presented in Fig.~\ref{fig:stability_analysis}(c), is
\begin{equation}
\begin{split}
F_k = \frac{{\rm k}_r P_x}{N}\Big( (F_r - e^{-Ts}F_k)&(K_p + K_i \frac{1}{s}) + F_r \\
&- K_{d,m} e^{-Ts}s N_m \frac{F_k}{{\rm k}_r} \Big).
\end{split}
\end{equation}
Then it becomes
\begin{equation}
\label{eq:pid_transfer}
\frac{F_k}{F_r} = \frac{{\rm k}_r P_x(K_p + K_i/s + 1)/N}{1 + e^{-Ts}P_x({\rm k}_r (K_p + K_i/s) +  K_d s N_m)/N}.
\end{equation}
When we apply a DOB instead of integral control, we need the inverse of the plant. In our case, the plant of the DOB is ${\rm PD_m}$, which is similar to Eq.~\eqref{eq:pid_transfer} except that $K_i$ and $e^{-Ts}$ are omitted:
\begin{equation}
P_{\rm{PD}_m} (= P_c) = \frac{{\rm k}_r P_x(K_p + 1)}{N + P_x({\rm k}_r K_p +  K_{d,m} s N_m)}.
\end{equation}
The formulation of $\rm{PD}_m$ including the DOB, which is shown in Fig.~\ref{fig:stability_analysis}(c), is 
\begin{equation}
F_k = \frac{{\rm k}_r P_x(K_p + 1)(F_d - e^{-Ts}P_c^{-1}Q_{\tau d} F_k)}{\left(N + e^{-Ts}P_x({\rm k}_r K_p + K_{d, m} s N_m)\right)\left(1-Q_{\tau d}\right)},
\end{equation}
where $Q_{\tau d}$ is a second order low-pass filter. Then the transfer function is 
\begin{equation}
\frac{F_k}{F_d} = \frac{{\rm k}_r P_x (K_p + 1)}{N(1-Q_{\tau d}) + e^{-Ts}\left(NQ_{\tau d} + P_x ({\rm k}_r K_p + K_{d,m} s N_m)\right)}.
\end{equation}
The open-loop transfer function is 
\begin{equation}
P_{\rm{PD}_m + \rm{DOB}}^{open} = \frac{NQ_{\tau d} + P_x ({\rm k}_r K_p + K_{d,m} s N_m)}{N(1-Q_{\tau d})}
\end{equation}

The bode plots of $P_{\rm{PD}_f}^{open}$, $P_{\rm{PD}_m}^{open}$, $P_{\rm{PID}_m}^{open}$, and $P_{\rm{PD}_m + DOB}^{open}$ are presented in Fig.~\ref{fig:stability_analysis}(b). The gains ($K_p$, $K_{d,m}$, $K_i$) are the same as the values that we use in the experiments presented in Section~\ref{sec:single_act_test}, which are 4, 15, and 300, respectively. The $\rm{PD}_f$ controller uses $K_d N_m / {\rm k}_r$ for $K_{d,f}$ to normalize the derivative gain. The cutoff frequency of the DOB is set to 15\si{\hertz} because this is where the $\rm{PD}_m + \rm{DOB}$ shows a magnitude trend similar to the integral controller ($\rm{PID}_m$). The results imply that the $\rm{PD}_m + \rm{DOB}$ controller is more stable than $\rm{PID}_m$ with respect to phase margin and maximum phase lag. This analysis is also experimentally verified in Section~\ref{sec:single_act_test}.


\section{Robotic Testbed}
\label{sec:singleleg_testbed}
\begin{figure}
\centering
\includegraphics[width=0.95\columnwidth]{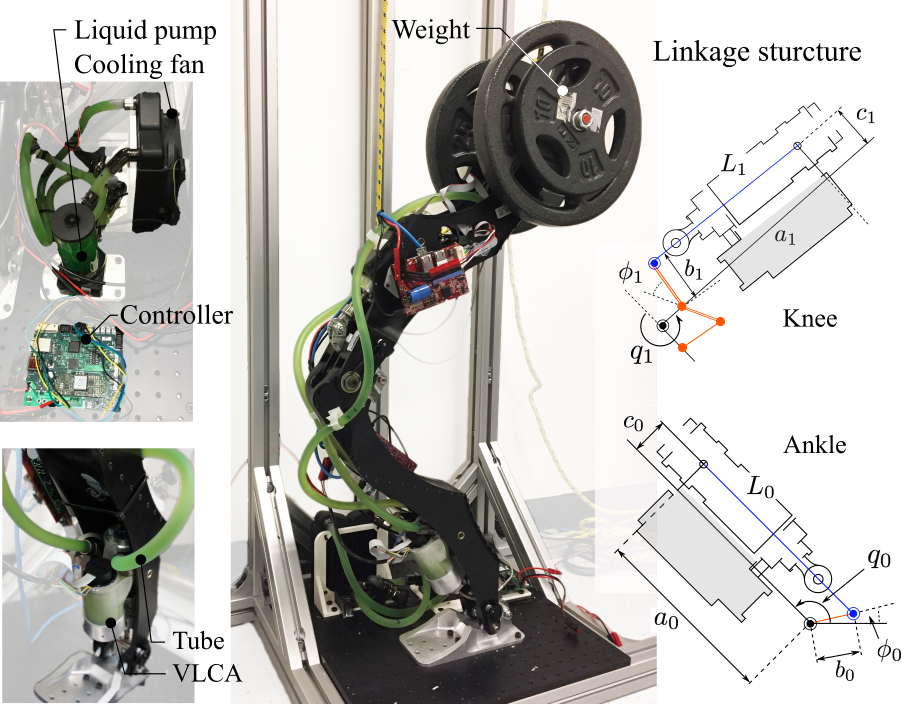}
\caption{ {\bf Robotic testbed.} Our testbed consists of two VLCAs at the ankle and the knee. The foot of the testbed is fixed on the ground.  The linkages are designed to vary the maximum peak torques and velocities depending on postures. As the joint positions change, the ratios between ball screw velocities ($\dot{L}_{0, 1}$) and joint velocities ($\dot{q}_{0,1}$) also change because of effective lengths of moment arms vary. The linkages are designed to exert more torque when the robot crouches, which is the posture that the gravitational loads on the joints are large.}
\label{fig:single_leg_testbed}
\vspace{-3mm}
\end{figure}

We built a robotic testbed shown in Fig.~\ref{fig:single_leg_testbed}. To demonstrate dynamic motion, we implemented an operational space controller (OSC) incorporating the multi-body dynamics of the robot.
%
%
We designed and built a robotic testbed (Fig.~\ref{fig:single_leg_testbed}) consisting of two VLCAs \-- one for the ankle ($q_0$) and one for the knee ($q_1$). The design constrains motion to the sagittal plane, the robot carries 10\si{\kilo\gram}, 23\si{\kilo\gram}, or 32.5\si{\kilo\gram} of weight at the hip, and the foot is fixed on the ground. With this testbed, we intended to demonstrate coordinated position control with two VLCAs, the viability of liquid cooling on an articulated platform, cartesian position control of a weighted end effector, and verification of a linkage design. 

The two joints each have a different linkage structure that was carefully designed so that the moment arm accommodates the expected torques and joint velocities as the robot posture changes (Fig.~\ref{fig:single_leg_testbed}). For example, each joint can exert a peak torque of approximately 270 $\si{\newton\metre}$ and the maximum joint velocity ranges between 7.5 $\si{\radian\per\second}$ and 20+ $\si{\radian\per\second}$ depending on the mechanical advantage of the linkage along the configurations. The joints can exert a maximum continuous torque of 91 $\si{\newton\metre}$ at the point of highest mechanical advantage. This posture dependent ratio of torque and velocity is a unique benefit of prismatic actuators.

Given cartesian motion trajectories, which are 2nd order B-spline or sinusoidal functions, the centralized controller computes the torque commands with operational space position and velocity, which are updated by the sensed joint position and velocity. The OSC formulation that we use is
\begin{equation}\label{eq:dyn_eq}
\bm{\tau} = \bm{A}{\bm{J}}_{hip}^{-1} (\ddot{\mathbf{x}}^{des} + K_p \mathbf{e} + K_d \dot{\mathbf{e}} - \dot{\bm{J}}_{hip}\dot{\mathbf{q}}) + \mathbf{b} + \mathbf{g},
\end{equation}
where $\bm{A}$, $\mathbf{b}$, and $\mathbf{g}$ represent inertia, coriolis, and gravity joint torque, respectively. $\ddot{\mathbf{x}}^{des}$, $\mathbf{e}$, and $\dot{\mathbf{e}}$ are desired trajectory acceleration, position and velocity error, respectively.
$\dot{\mathbf{q}} \in \mathcal{R}^2$ is the joint velocity of the robot and $\mathbf{\tau}$ is the joint torque. $\bm{J}_{hip}$ is a jacobian of the hip, which is a $2\times2$ square matrix and assumed to be full-rank.



\begin{figure*}
\centering
\includegraphics[width=2.0\columnwidth]{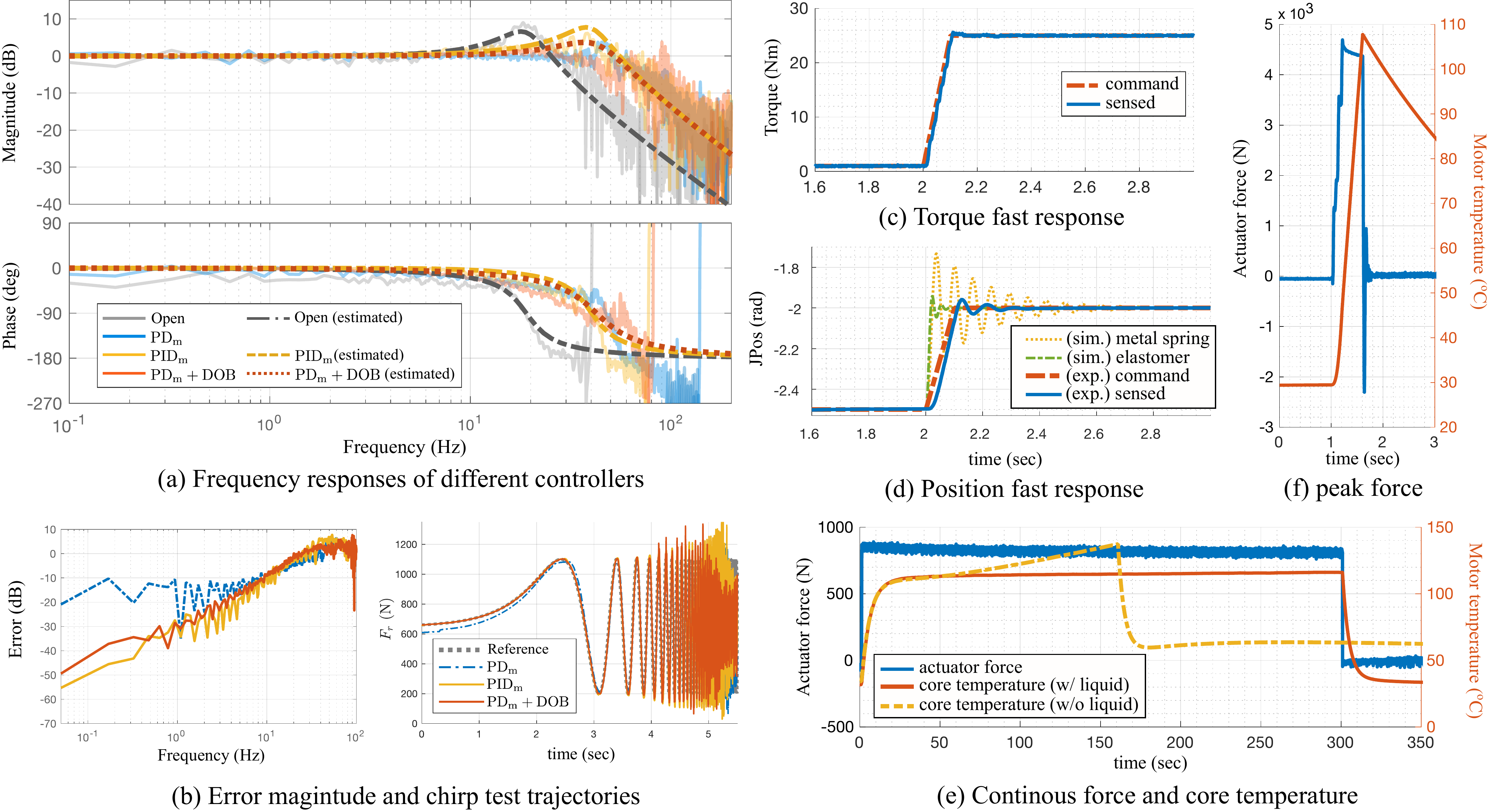}
\caption{{\bf Torque Feedback Control Test.} (a) Experimental data and estimated response based on the transfer functions are presented. Estimated response of PD controller is identical to the PD+DOB since DOB theoretically does not change the transfer function. The plot show PD+DOB shows better performance in terms of less overshoot and smaller phase drop near to the natural frequency. (b) We choose integral controller feedback gain that shows similar accuracy of PD+DOB's. The left is error magnitude of three controllers. PD controller has larger error than the other two controller in the low frequency region. The right is torque trajectories in the time domain.} 
\label{fig:actuator_test}
\vspace{-2.5mm}
\end{figure*}

\section{Results}
\label{sec:exp_result}

We first conducted various single actuator tests to show basic performance such as torque and joint position controllability, continuous and peak torque, and impact resistance. Subsequently, we focused on the performance of OSC using the robotic testbed integrated with DOB based torque controllers to demonstrate actuator efficiency and high power motions. 

\subsection{Single Actuator Tests}
\label{sec:single_act_test}
\Copy{single_actuator_test}{

Fig.~\ref{fig:actuator_test}(a) shows the experimental results of our frequency response testing as well as the estimated response based on the transfer functions. We compare three types of controllers: $\rm{PD}_m$, $\rm{PID}_m$, and $\rm{PD}_m + \rm{DOB}$. As we predicted in the analysis of Section~\ref{sec:act_force_ctrl}, the $\rm{PD}_m + \rm{DOB}$ controller shows less phase drop and overshoot than $\rm{PID}_m$. The integral control feedback gain used in the experiment is 300 and the cutoff frequency of the DOB's $Q_{\tau d}$ filter is 60\si{\hertz}, which shows similar error to the $\rm{PID}_m$ controller (Fig.~\ref{fig:actuator_test}(b)). Another test presented in Fig.~\ref{fig:actuator_test}(c) also supports the stability and accuracy of torque control. In the test, we command a ramp in joint torque from 1 to 25\si{\newton\meter} in 0.1\si{\second}. The sensed torque (blue solid line) almost overlaps the commanded torque (red dashed line).

\Copy{jpos_ctrl}{
Fig.~\ref{fig:actuator_test}(d) is the result of a joint position control test designed to show that VLCAs have better joint position controllability than SEAs using springs. In the experiment, we use a joint encoder for position control and a motor quadrature encoder for velocity feedback. To compare the VLCAs performance with that of spring-based SEAs, we present simulation results for a spring-based SEA on the same plot as the experiment result for the VLCA. The green dashed line is the simulated step response of our actuator and the yellow dotted line is the result of the simulation model using the same parameters except the spring stiffness and damping. The spring stiffness was selected to be 11\% of the elastomer's, based on the results of our tests in Section~\ref{sec:visco}, and the damping for the spring case was set to 8000 \si{\newton \second \per \meter} which only includes the drivetrain friction. The results show a notable improvement in joint position control when using an elastomer instead of a steel spring.
}

Fig.~\ref{fig:actuator_test}(e) shows the continous force and the motor core temperature trend with and without liquid cooling. The observed continous force is 860\si{\newton} and the motor core temperature settles at 115\si{\degreeCelsius} with liquid cooling. 
Fig.~\ref{fig:actuator_test}(f) is the the result of short-term torque test. In the experiment, we fix the output of the actuator and command a 31\si{\ampere} current for 0.5\si{\second}. \Copy{max_torque_density}{The observed force measured by a loadcell (Fig.~\ref{fig:actuator}(c)) is 4500\si{\newton}, which is a little smaller than the theoretically expected value, 5900\si{\newton}. Considering that the estimated core temperature surpassed 107\si{\degreeCelsius} ($< 155\si{\degreeCelsius}$ limit), we expect that the theoretical value is reasonable. Thus, we conclude that the maximum force density of our actuator is larger than 2700\si{\newton\per\kilo\gram} and potentially 3500\si{\newton\per\kilo\gram}.}
}

\begin{figure}
\centering
\includegraphics[width=\columnwidth]{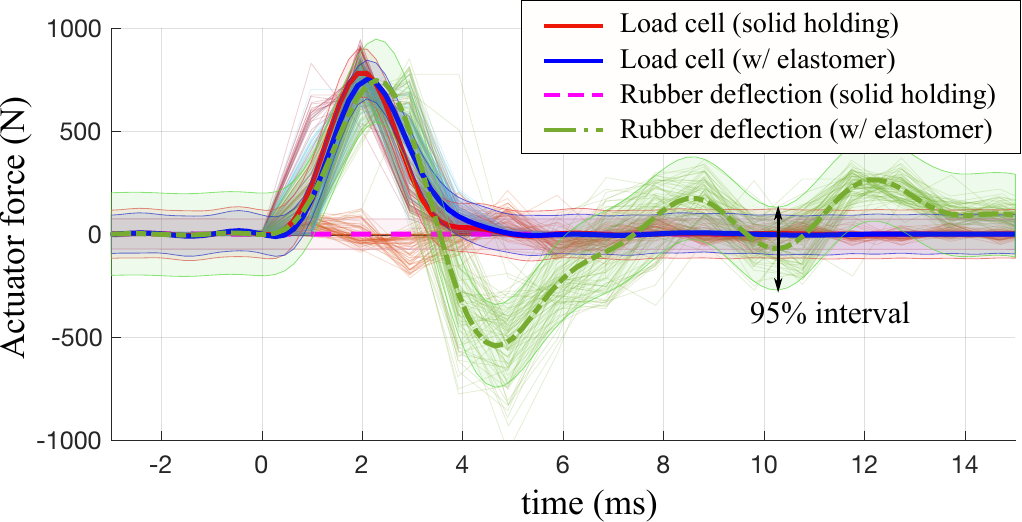}
\caption{{\bf Impact test.} 83 trials are plotted and estimated with gaussian process. We can see the deflections of the elastomer, which imply that the elastic element absorbes the external impact force. }
\label{fig:impact_test}
\vspace{-2.5mm}
\end{figure}

\Copy{impact_test}{
Fig.~\ref{fig:impact_test} shows loadcell and elastomer force data from the impact tests. In the tests, we hit the loadcell connected to the ball screw (Fig.~\ref{fig:actuator}(c)) with a hammer falling from a constant height while fixing the actuator in two different places to compare the rigid actuator to viscoelastic actuator response. In the rigid scenario, outer case of ballnut, a blue part in Fig.~\ref{fig:actuator}, is fixed to exclude the elastomer from the external impact force path. In the second case, we fixed the ground pin of the actuator, which is depicted by a gray part in Fig.~\ref{fig:actuator}(l), to see how the elastomers react to the impact.

The impact experiment is challenging because the number of data points we can obtain is very small with a 1\si{\milli\second} update rate. To overcome the lack of data points, we estimate the mean and variance of 83 trials by gaussian process regression. The results presented in Fig.~\ref{fig:impact_test} imply that there is no significant difference in the forces measured by the loadcell in both cases, which is predictable because the elastic element is placed behind the drivetrain. However, the elastomer does play a significant role in absorbing energy from the impact which is evident from large elastomer deflection in the second case. Thus, the presence of the elastic element mitigates the propagation of an impulse to the link where the actuator grounds.
}

\begin{figure}
\centering
\includegraphics[width=\columnwidth]{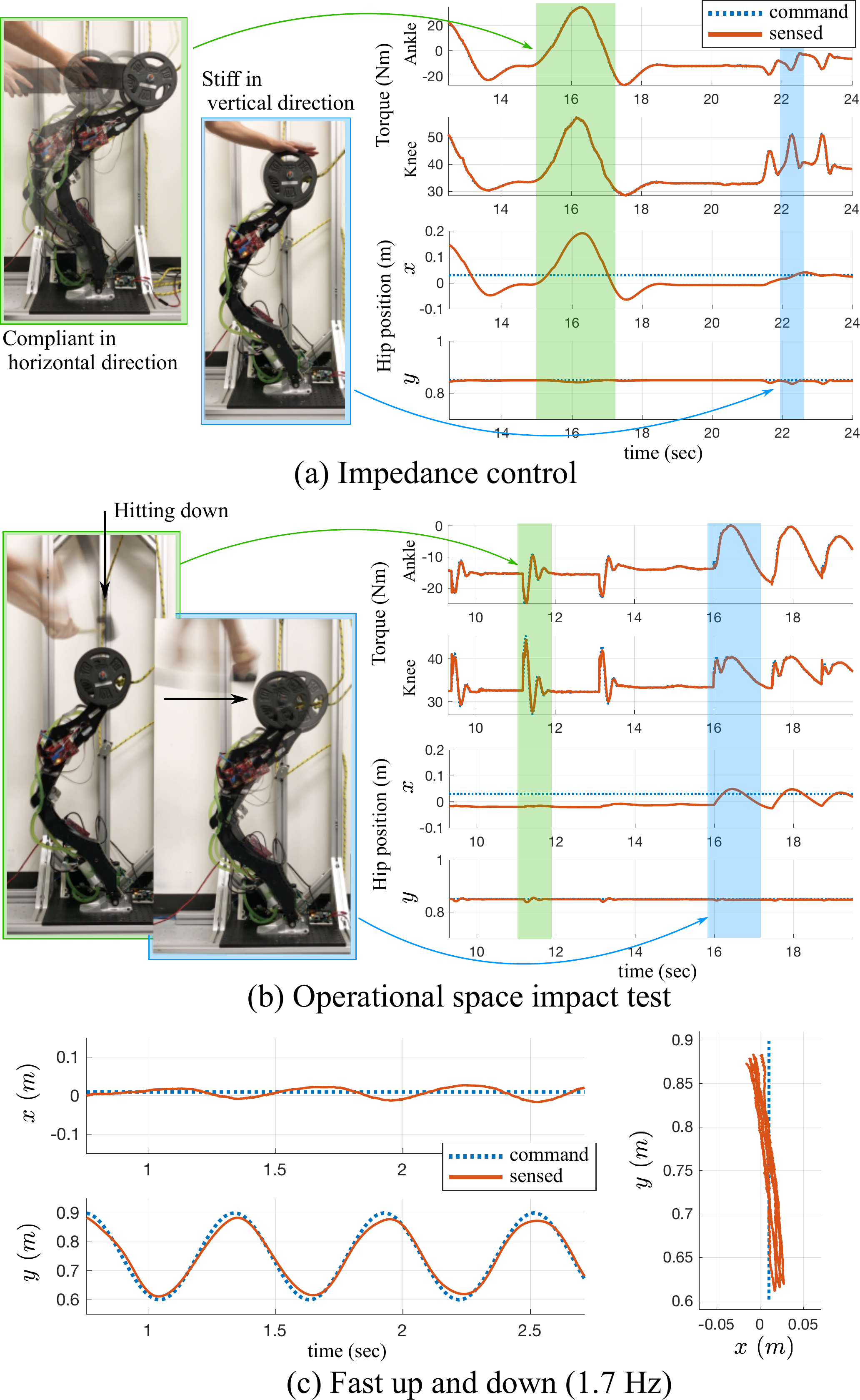}
\caption{{\bf Operational Space Impedance Control Test.} (a) The robot demonstrates different impedance: stiff in the vertical direction and compliant in the horizontal direction. The high tracking performance of force feedback control results in the overlapped commanded and sensed torques. (b) To show the stability, we hit the weight with a hammer while operating the impedance control. Even under the impact, force control show stable and accurate tracking. (c) The robot demonstrates a 1.7\si{\hertz} up and down motion while carrying 10\si{\kilo\gram} weight at the hip, and shows a position error of less than 2.5\si{\centi\meter}.}
\label{fig:impedance_ctrl}
\end{figure}

\begin{figure}
\centering
\includegraphics[width=0.9\columnwidth]{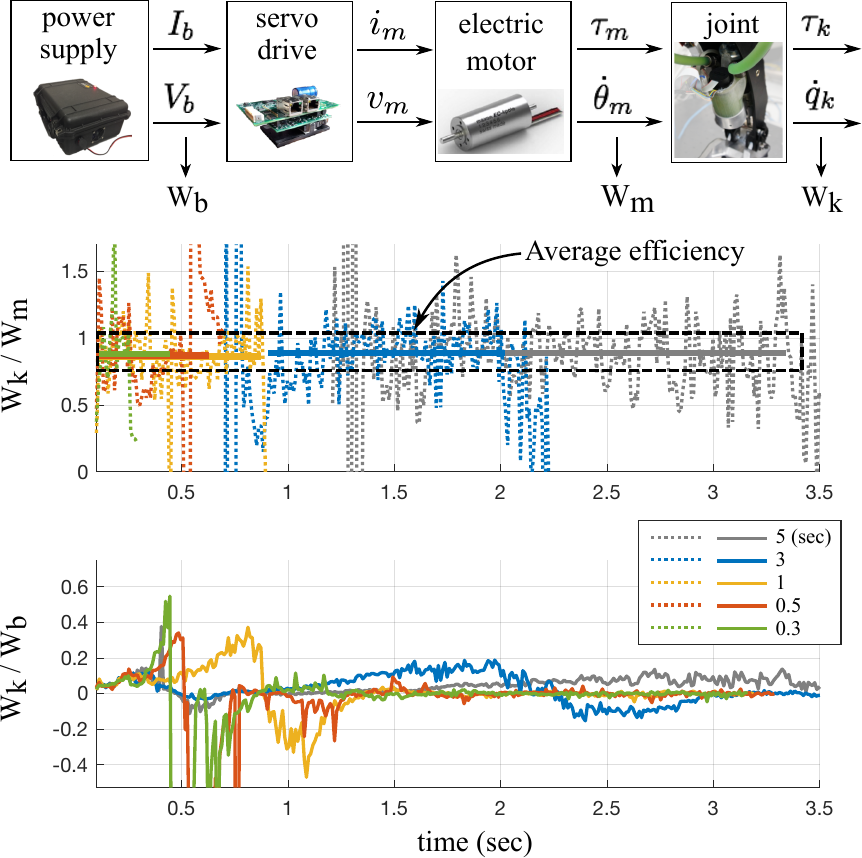}
\caption{{\bf Efficiency analysis of the ankle actuator.} Efficiencies of mechanical system using electrical power has 3 steps from a power supply to robot joints. The graph shows the ratio of the mechanical power of the ankle joint and the motor power and the ratio of the joint power and power supply's input power.}
\label{fig:efficiency}
\vspace{-2.5mm}
\end{figure}

\subsection{Operational Space Impedance Control}
\label{sec:osc_test}

Fig.~\ref{fig:impedance_ctrl} shows our OSC experimental tests (Section~\ref{sec:singleleg_testbed}) carrying a 10\si{\kilo\gram} weight. In the first test presented in Fig.~\ref{fig:impedance_ctrl}(a), the commanded behavior is to be compliant in the horizontal direction ($x$) and to be stiff in the vertical direction ($y$). When pushing the hip with a sponge in the $x$ direction, the robot smoothly moves back to comply with the push, but it strongly resists the given vertical disturbance to maintain the commanded height. To show the stability of our controller, we also test the response to impacts by hitting the weight with a hammer (Fig.~\ref{fig:impedance_ctrl}(b)). Even when there are sudden disturbances, the torque controllers rapidly respond to maintain good torque tracking performance as shown in Fig.~\ref{fig:actuator_test}(d).

\Copy{osc_updown}{
Fig.~\ref{fig:impedance_ctrl}(c) shows the tracking performance of our system while following a fast vertical hip trajectory. While traveling 0.3\si{\meter} with 1.7\si{\hertz} frequency, the hip position errors are bounded by 0.025\si{\meter}. This result demonstrates that our system is capable of stable and accurate OSC, which is challenging because of the bandwidth conflict induced by its cascaded structure. 
}

\subsection{Efficiency Analysis}
\label{sec:efficiency_analysis}
\begin{figure*}
\centering
\includegraphics[width=1.9\columnwidth]{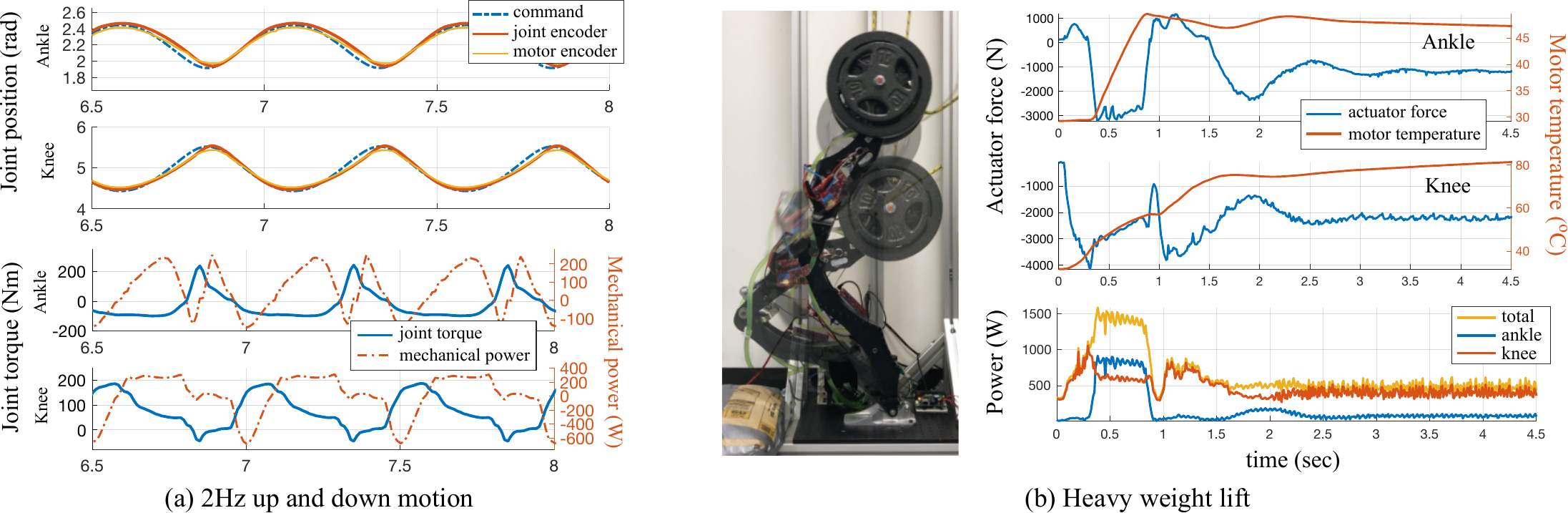}
\caption{{\bf High power motion experiment.} (a) Joint position data from joint encoder and motor encoder are shown. In this experiment, the maximum observed torque of the ankle joint is 250 \si{\newton\meter} and the maximum observed mechanical power of the knee joint is 310\si{\watt}. (b) The robot lifts by 0.3\si{\meter} a 32.5 \si{\kilo\gram} load during 0.4\si{\second}. There is still a safety margin with respect to the limits equal to 5900\si{\newton} and 155\si{\degreeCelsius}.}
\label{fig:high_power_test}
\vspace{-2.5mm}
\end{figure*}

Fig.~\ref{fig:efficiency} explains the power flow from the power supply to the robot joint. Input current ($I_b$) and voltage ($V_b$) are measured in the micro-controllers and the product of those two yields the input power from the power supply. $\dot{\theta}_m$ is measured by the quadrature encoder connected to the motor's axis (Fig.~\ref{fig:actuator}(f)) and $\tau_m$ is computed from $k_{\tau}i_m$ with $i_m$ measured in the micro-controller. \Copy{joint_encoder_velocity}{Joint velocity is low-pass derivative filtered joint positions measured at the absolute joint encoders.} The torque ($\tau_k$) is computed from projecting the load cell data across the linkage's effective moment arm. 

In this test, the robot lifts a 23\si{\kilo\gram} load using five different durations to observe efficiency over a range of different speeds and torques. The results are presented in Fig.~\ref{fig:efficiency} with the description of three different power measures. The sensed torque data measured by a load cell is noisy; therefore, we compute the average of the drivetrain efficiency for a clearer comparison. The averages are the integrations of efficiency divided by the time durations. Here we only integrate efficiency while the mechanical power is positive, to prevent confounding our results by incorporating the work done by gravity. 

The experimental results show that the drivetrain efficiency is approximately 0.89, which means that we lose only a small amount of power in the drivetrain and most of the torque from the motor is delivered to the joint. This high efficiency indicates only minor drivetrain friction, which is beneficial for dynamics-based motion controllers.

\subsection{High Power Motion Experiment}
To demonstrate high power motions such as fast vertical trajectories and heavy payload lifts, we use the motor position control mode, which uses the quadrature encoders attached directly to the motor for feedback. Fig.~\ref{fig:high_power_test}(a) presents the results of a test comprised of 2\si{\hertz} vertical motion with 0.32 \si{\meter} of travel while carrying a load of 10 \si{\kilo\gram} at the hip. 
With respect to mechanical power, the knee joint repeatedly exerts 305\si{\watt}, which is close to the predicted constant power (360\si{\watt}). Although the limited range of motion makes it hard to demonstrate continuous mechanical power, these results convincingly support our claim of enhanced continuous power enabled through liquid cooling.

Fig.~\ref{fig:high_power_test}(b) presents another test in which the robot lifts a 32.5\si{\kilo\gram} weight. We can see that the robot operates in the safe region ($\leq$ 5900\si{\newton} and $\leq$ 155\si{\degreeCelsius}) while demonstrating high power motion.

\section{Concluding Remarks}
\label{sec:conclusion}
Overall our main contribution has been on the design and extensive testing of a new viscoelastic liquid cooled actuator for robotics. 

\Copy{osc-comment}{One of the tests addressed is impedance control in the operational space instead of joint impedance control. It is often the case that humanoid robots require impedance control in the operational space. For instance, controlling the operational space impedance can enable improved locomotion behaviors such as running. Our controllers demonstrate that we can control the impedance in the Cartesian operational space as a potential functionality for future robotic systems.} The use of liquid cooling has allowed to sustain high output torque for prolonged times as shown in the experiments of Fig. 6(e). As we can see, when turning off liquid cooling the temperature rises quickly above safety limits whereas when turning on the cooling we can sustain large payload torques for long periods of time. The use of elastomers versus steel springs has demonstrated a clear improvement on joint position performance as shown in Fig. 6(d). This capability is important to achieve a large range of output joint or Cartesian space impedances.


In the future we will explore further reducing the size of our viscoelastic liquid cooled actuators. Maintaining the current compact design structure we can still reduce another significant percentage the bulk of the actuator by exploring new types of bearings, ballnut sizes and piston bearings at the front end of the actuator. We will also explore using different material for the liquid cooling actuator jacket. The current polyoxymethylene material is easily breakable and develops cracks due to the vibrations and impacts of this kind of robotic applications. In the future we will switch to sealed metal chambers for instance. Further in the future we will consider designing our own motor stators and rotors for improved performance. We expect this kind of actuators to make their way into full humanoid robots and high performance exoskeleton devices and we look forward to participate in such interesting future studies.




\appendices

\section*{Acknowledgment}
The authors would like to thank the members of the Human Centered Robotics Laboratory at The University of Texas at Austin for their help and support. This work was supported by the Office of Naval Research, ONR Grant [grant \#N000141512507] and NASA Johnson Space Center, NSF/NASA NRI Grant [grant \#NNX12AM03G].

\ifCLASSOPTIONcaptionsoff
  \newpage
\fi

\bibliographystyle{IEEEtran}
\bibliography{ms}

\end{document}